# Tunable perpendicular magnetic anisotropy in GdFeCo amorphous films

Manli Ding[a)] and S. Joseph Poon[a)]

*Department of physics, University of Virginia, Charlottesville, Virginia 22904, USA*

**Abstract**

We report the compositional and temperature dependence of magnetic compensation in amorphous ferrimagnetic $Gd_xFe_{93-x}Co_7$ alloy films. Magnetic compensation is attributed to the competition between antiferromagnetic coupling of rare-earth (RE) with transition-metal (TM) ions and ferromagnetic interaction between the TM ions. The low-Gd region of x between 20 and 34 was found to exhibit compensation phenomena characterized by a low saturation magnetization and perpendicular magnetic anisotropy (PMA) near the compensation temperature. Compensation temperature was not observed in previously unreported high-Gd region of x=52-59, in qualitative agreement with results from recent model calculations. However, low magnetization was achieved at room temperature, accompanied by a large PMA with coercivity reaching ~6.6 kOe. The observed perpendicular magnetic anisotropy of amorphous GdFeCo films probably has a structural origin consistent with certain aspects of the atomic-scale anisotropy. Our findings have broadened the composition range of transition metal-rare earth alloys for designing PMA films, making it attractive for tunable magnetic anisotropy in nanoscale devices.

[a)]Electronic mail: md3jx@virginia.edu, sjp9x@virginia.edu.

# 1. Introduction

Magnetic materials with perpendicular magnetic anisotropy (PMA) have attracted large interest over the past few years from the viewpoint of both academic research and technological applications. It is predicted that magnetic tunnel junctions or spin valves with perpendicularly magnetized electrodes are able to facilitate faster and smaller data-storage magnetic bits, compared with the normal in-plane ones [1]. For devices, it is preferable to use a single film layer with perpendicular magnetic anisotropy instead of multilayer in order to avoid complicated fabrication process and decrease the total thickness of the devices. One well-known candidate is amorphous rare-earth transition-metal (RE–TM) thin films with strong perpendicular magnetic anisotropy [2,3,4].

Amorphous GdFeCo films have been known to have a low saturation magnetization, preventing magnetization curling at the film edge. Furthermore, they possess Curie temperatures well above room temperature for a wide range of compositions [5]. Gd is a unique member of the lanthanide series in that its ground-state electronic configuration is $4f^{7}(5d6s)^{3}$ with the highest possible number of majority spin electrons and no minority spin electron in its $4f$ state according to Hund's rule. In addition, because the $4f$ states of Gd are half filled, their orbital moment and spin-orbit coupling are zero. This L=0 state of Gd provides a favorable condition for low Gilbert damping, which is preferable in spin-torque-transfer devices. Amorphous GdFeCo alloys are ferrimagnets in which the Fe(Co) sublattices are antiferromagnetically coupled to the Gd sublattice in a collinear alignment, while the exchange coupling in the Fe(Co) sublattice is ferromagnetic [6]. These ferrimagnetic GdFeCo alloys tend to exhibit magnetic compensation behavior characterized by a vanishing magnetization below the Curie temperature [7]. Also, depending on the composition, amorphous GdFeCo films generally possess a uniaxial anisotropy with an anisotropy axis either perpendicular or parallel to the film plane.

In this study, we have obtained amorphous GdFeCo films for a wide range of Gd content via the combinatorial growth technique. We have investigated the compositional and temperature dependence of magnetization compensation in these amorphous ferrimagnetic $Gd_xFe_{93-x}Co_7$ films and demonstrated the tunability of perpendicular magnetic anisotropy. Possible mechanisms for the observed perpendicular magnetic anisotropy are discussed.

## 2. Experimental procedure

The $Gd_xFe_{93-x}Co_7$ (GdFeCo) films were prepared at ambient temperature on thermally oxidized Si substrates using *rf* magnetron sputtering. The base pressure of the sputtering chamber was ~ $7 \times 10^{-7}$ Torr. GdFeCo alloy films were deposited by means of co-sputtering with the elemental targets under a processing Ar gas pressure around $5 \times 10^{-3}$ Torr. The capping MgO layer was formed directly from a sintered MgO target to protect GdFeCo layer from oxidation. All the samples deposited at room temperature had a typical structure consisting of Si(100)/SiO$_2$/Gd$_x$Fe$_{93-x}$Co$_7$($15 \leq x \leq 59$ at. %)/MgO(6 nm) with a fixed thickness of GdFeCo layer ~ 50 nm. The film thickness of all samples were measured by x-ray reflectivity, and film compositions were determined using inductively coupled plasma-mass spectrometry (ICP-MS) after chemically dissolving the films, and confirmed by X-ray fluorescence (XRF) using peak ratios. The magnetic properties of the samples were investigated by vibrating sample magnetometer (VSM) and magneto-optic Kerr effect (MOKE) measurements, with a maximum field of 20 kOe. Structural characterization of the films were performed by x-ray diffraction (XRD) with Cu *K*α ($\lambda = 1.541$ Å) radiation (Smart-lab®, Rigaku Inc.) and transmission electron microscopy (TEM, FEI Titan). Atomic Force Microscopy (Cypher™, Asylum Research Inc.) was used to characterize surface morphology.

## 3. Results and discussions

The amorphous structures of the as-deposited samples were confirmed by TEM and XRD observations. Fig. 1(a) shows a typical cross-sectional TEM image obtained from as-deposited $Gd_{22}Fe_{71}Co_7$ film on the Si/ $SiO_2$ substrate. The microstructure was dense with no visible cracks or holes, and all layers in the structure were well adhered to each other. The uniform thickness of $Gd_{22}Fe_{71}Co_7$ film was measured to be 52 nm in Fig. 1(a). The high resolution TEM image in Fig. 1(b) revealed the featureless nanoscale structure that indicated the lack of long-range order. The broad ring pattern in the fast Fourier-transform (FFT) image in the inset of Fig. 1(b) indicated the lack of crystallinity. In addition, XRD scans showed that there were no diffraction peaks other than those from the substrate for GdFeCo films with different compositions (not shown here). The XRD results were consistent with the TEM, indicating that all the as-deposited samples were amorphous in nature, without the formation of a long-range structural order. Atomic force microscopy (AFM) measurements showed that the surfaces were free of pinholes and were flat with roughness less than 1 nm.

### 3.1. Low Gd-content films

The magnetization of GdFeCo films were characterized in the in-plane and out-of-plane directions using the VSM option in Quantum Design VersaLab. Fig. 2 shows the temperature dependence of saturation magnetization of as-deposited GdFeCo films for several compositions. Saturation magnetization ($M_s$) was extracted from the hysteresis loops measured as a function of temperature between 100 K and 400 K. Compensation temperature ($T_{comp}$) was defined as the temperature at which $M_s$(T) reached its minimum. The saturation moments at the compensation temperatures were below 100 emu/cc. The observed small saturation moment was due to the ferrimagnetism of amorphous RE-TM alloys. The TM-TM ferromagnetic interaction aligns the magnetic moments among Fe and Co ions, which are coupled antiferromagnetically with the magnetic moments of Gd. As a result, the net moment is the difference between the magnetic moments of Gd and Co(Fe). At the compensation temperature, the moments of the two magnetic sublattices were nearly equal, giving rise to a low saturation magnetization.

Due to the different temperature dependence of the sublattice magnetizations, compensation temperature can be varied, depending on the compositions. The variation of $M_s$ with temperature (Fig. 2) resembled the expected compensation behavior when approaching the compensation point. The GdFeCo films with x= 22, 27 and 30 at. % exhibited the compensation temperatures $T_{comp}$ at 300 K, 350 K and 378 K, respectively. However, for several other compositions, there was no compensation point within the investigated temperature interval from 100 to 400 K. For $Gd_{35}Fe_{58}Co_7$ film, the compensation point was not obtained due to the limitation of the measurement temperature range. For the sample $Gd_{15}Fe_{78}Co_7$, the dependence of saturation magnetization on $T$ (not shown) indicated that the magnetization of the Fe(Co) sublattices exceeded the magnetization of the Gd sublattice in the whole temperature range. This was due to the fact that ferromagnetic exchange of the Fe(Co) sublattices dominates the magnetic behavior at low Gd content. As the Gd content increases and Fe(Co) content decreases, there is a corresponding increase in the antiferromagnetic coupling relative to the ferromagnetic exchange, and magnetic compensation emerges. Further increase in the Gd concentration resulted in the increase in $T_{comp}$, as shown in the inset to Fig. 2.

According to magnetization measurements, the perpendicular magnetic anisotropy in GdFeCo films appeared near their compensation temperatures, whereas otherwise the magnetic easy axis was in-plane. Fig. 3(a) shows the typical normalized out-of-plane hysteresis loops of the as-deposited $Gd_{27}Fe_{66}Co_7$ films at various temperatures. At 250K, far from $T_{comp}$~350K, magnetization was dominated by in-plane anisotropy. Near the compensation point, perpendicular magnetic anisotropy was dominant, and a square hysteresis loop was established in the out-of-plane direction with coercivity near 100 Oe.

From the magnetization variation curve (Fig. 2), the $Gd_{22}Fe_{71}Co_7$ film had magnetic compensation temperature at room temperature, which has particular technological importance [8]. Fig. 3(b) shows the room-temperature in-plane and out-of-plane hysteresis (M-H) loops of the 50 nm $Gd_{22}Fe_{71}Co_7$ film, indicating that this composition has PMA with out-of-plane coercivity of about 360 Oe. The total perpendicular anisotropy energy density ($K_u$), which determines the thermal stability, was $3.8 \times 10^5$ erg/cm$^3$, as calculated by evaluating the area enclosed between the in-plane and perpendicular M-H curves [9]. The out-of-plane loop showed sharp, square switching characteristics with a squareness of one (1). For $Gd_{22}Fe_{71}Co_7$ film, a relatively small concentration of the high-moment RE magnetically compensated the lower moment TM at room temperature. Since the tunneling current in spintronics devices is dominated by that from the highly polarized TM atoms, this material would be very useful for device applications because it would eliminate magnetic dipole fields that can give rise to significant magnetic coupling within and between devices.

The above results underlined the fact that in GdFeCo system, the magnetization anisotropy may be easily tuned by adjusting the composition and/or the temperature. In the low-Gd region of x between 20 and 34, the Gd sublattice dominated the overall magnetization of the system below the compensation temperature. However, Above $T_{comp}$ the magnetization of the Fe(Co) sublattice prevailed and the total magnetization of the system continuously increased on further warming. Since the magnetic anisotropy can vary around this temperature, $T_{comp}$ is an important parameter that ensures the stability of the stored information.

### 3.2. High Gd-content films

At room temperature, amorphous GdFeCo films with Gd concentrations varying between 52% and 59% were found to exhibit low magnetization. Fig. 4 shows the temperature dependence of saturation magnetization of as-deposited GdFeCo films for Gd concentrations at 54% and 57%. The magnetization decreased precipitously with increasing temperature and finally decreased to near zero between 370 and 400 K. Magnetic compensation temperature was not observed in this composition region, which was consistent with some recently reported computational results on GdFeCo films [10]. The film compositions with 54 and 57 at. % Gd exhibited Curie temperatures at 375 K and 400 K, respectively, much lower than 500 K of the composition with 22 at. % Gd [11]. These can be attributed to a stronger role of antiferromagnetic coupling at high Gd content, which also tended to reduce the Curie temperature.

In view of the low magnetization, perpendicular magnetic anisotropy was investigated at room temperature for Gd content between 52% to 59%. Fig. 5(a) shows the compositional dependence of out-of-plane coercivity ($H_c$) and saturation magnetization ($M_s$) of 50 nm as-deposited perpendicularly magnetized GdFeCo films for x between 52 and 59. With the increasing content of Gd, $M_s$ decreased from 100 to 55 emu/cc at x=57, then increased to 84 emu/cc at x=59. However, the out-of-plane coercivity showed an opposite trend, which increased with increasing Gd concentration and was greatest in the sample with 57 at. % Gd. Fig. 5(b) shows the typical hysteresis loops of as-deposited $Gd_{57}Fe_{36}Co_7$ film with thickness of 50 nm. A clear perpendicular anisotropy was realized with out-of-plane coercivity $H_c = 6.6$ kOe. The total perpendicular anisotropy energy density $K_u$, which determined the thermal stability, was $2.6 \times 10^5$ erg/cc. Generally, the RE ions exhibit large local magnetic anisotropy due to its spin-orbit coupling. Gd is in the L=0 state, spin-orbit coupling is supposed to be small; however, the 5d electrons have finite spin-orbit coupling which can be partially responsible for this anisotropy. Because of their large out-of-plane coercivity, ferrimagnetic GdFeCo films in the composition range $52 \leq x \leq 59$ can be of technological importance in the area of the thermomagnetic recording devices.

### 3.3. Discussion

Perpendicular magnetic anisotropy at room temperature was found in the compositions near 23 and 57 at. % Gd. Since amorphous alloys lack structural long-range order, one might expect that atomic-scale structure plays an important role in determining the properties of these alloys. Studies have attempted to correlate the magnetic anisotropy of amorphous RE-TM films with various structural characteristics ranging from columnar textures [12] to microcrystallinity [13] to local magnetic anisotropy or/and atomic-scale anisotropy [14,15,16]. Cross-sectional TEM study (Fig. 1) failed to detect nanoscale columnar growth and microcrystallinity in the amorphous GdFeCo films. As discussed above, local magnetic anisotropy of rare earth atoms does not play a role in Gd-Fe-Co, unlike other rare earth elements such as Tb and Dy; and yet amorphous Gd-Fe-Co films show robust PMA behavior with a large $H_c$ comparable to that of amorphous Tb-Fe-Co films [17]. These findings suggest that the observed perpendicular magnetic anisotropy in amorphous GdFeCo films probably has a structural origin involving atomic-scale anisotropy. This atomic-scale order can be anisotropic in as-grown films, which influences the short-range exchange interaction, leading to anisotropic properties. The perpendicular anisotropy may arise from the change in nearest neighbor distance and the coordination numbers for RE and TM sublattices which can affect the short range order.

## 4. Conclusion

In summary, it was shown that the magnetic anisotropy of amorphous ferrimagnetic $Gd_xFe_{93-x}Co_7$ films can be controlled by varying the composition as well as temperature. The low-Gd region of x between 20 and 34 were found to exhibit compensation phenomena characterized by a low saturation magnetization and perpendicular magnetic anisotropy near the compensation temperature. Furthermore, low magnetizations and perpendicular magnetic anisotropy with large coercivity of 6.6 kOe were observed at room temperature in previously unreported composition range of 52-59 at. % Gd. No compensation temperature was measured in this range, which was consistent with recent model calculations. The observed perpendicular magnetic anisotropy in amorphous GdFeCo films probably has a structural origin consistent with certain aspects of the atomic-scale anisotropy. Our results have provided a way to fabricate GdFeCo films with tunable magnetic anisotropy by varying the composition or/and temperature, making these amorphous films attractive for future nanomagnetic devices.

**Acknowledgements**

We thank Dr. T. Paul Adl in Micron Techology for helpful composition measurements and Prof. Jiwei Lu (University of Virginia) for stimulating discussions. This work was carried out under the financial support of DARPA through a subcontract with Grandis Inc.

Figure Captions:

**Fig. 1.** (a) Cross-sectional TEM image of as-deposited $Gd_{22}Fe_{71}Co_7$ film on $SiO_2$/Si substrate. 6 nm MgO layer was used to cap the film; (b) high-resolution TEM image of the $Gd_{22}Fe_{71}Co_7$ film. The inset is a FFT pattern of the image.

**Fig. 2.** Temperature dependence of the saturation magnetization of as-deposited GdFeCo films with various Gd concentrations (x = 22, 27, 30 and 35). The inset in (b) shows the dependence of the magnetization compensation temperature $T_{comp}$ on the Gd concentration.

**Fig. 3.** (a) Normalized out-of-plane hysteresis loops of as-deposited $Gd_{27}Fe_{66}Co_7$ film measured at 250, 300, and 350 K. (b) In-plane (black square) and out-of plane (red circle) hysteresis loops of 50 nm as-deposited $Gd_{22}Fe_{71}Co_7$ film.

**Fig. 4.** Temperature dependence of the saturation magnetization of GdFeCo films with various Gd concentrations (x = 54 and 57).

**Fig. 5.** (a) Gd-content dependence of $M_s$ (black triangle) and $H_c$ (blue circle) for perpendicularly magnetized GdFeCo films between 52 and 59 at. % Gd. (b) In-plane and out-of-plane hysteresis loops of $Gd_{57}Fe_{36}Co_7$ film with thickness at 50 nm.

Figure 1:

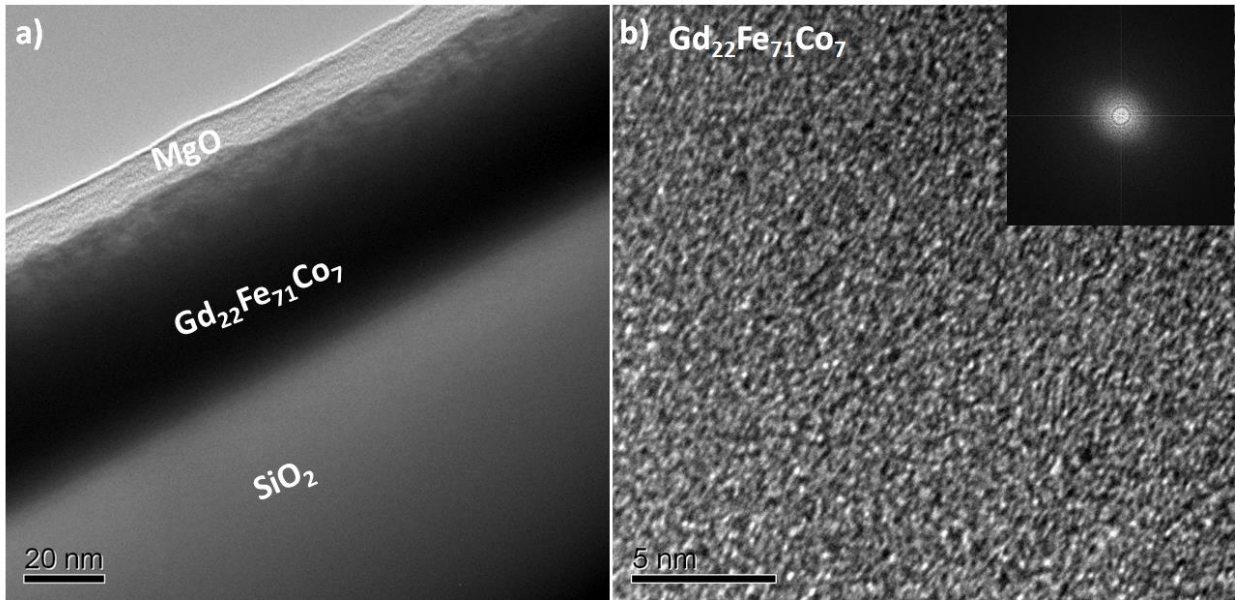

Figure 2:

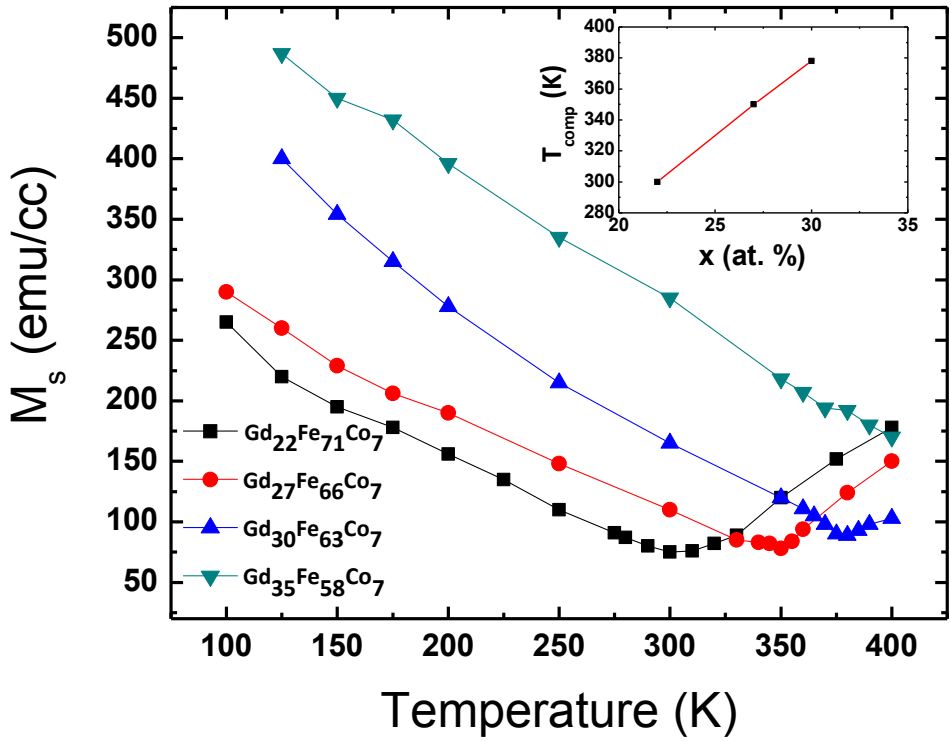

Figure 3:

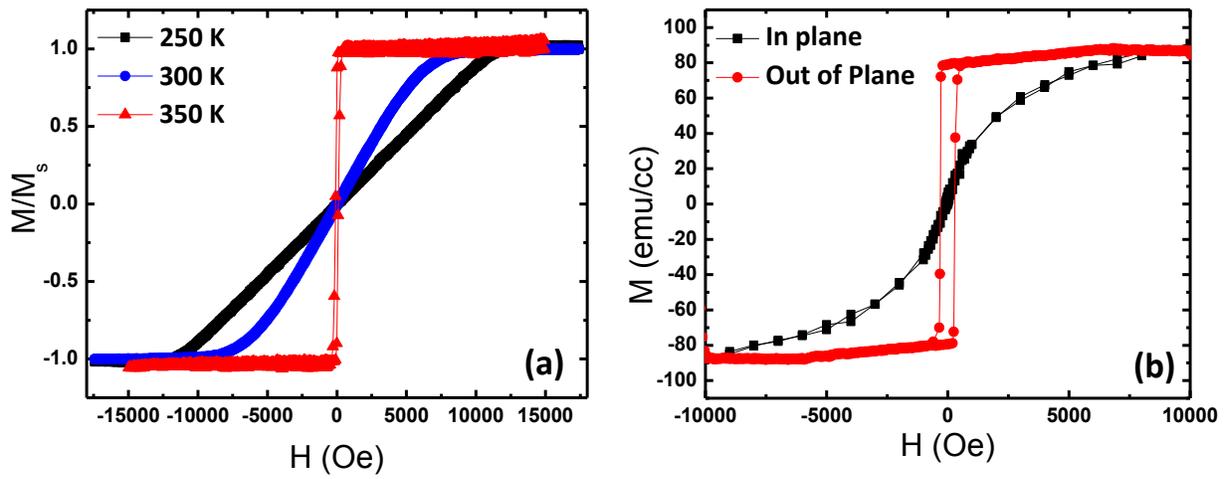

Figure 4:

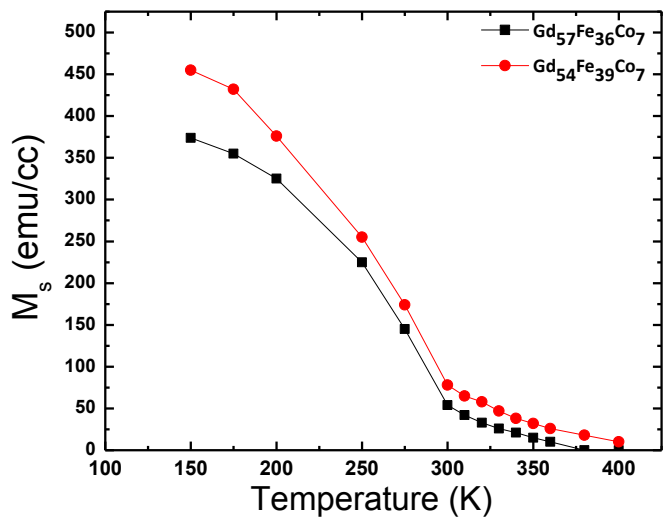

Figure 5:

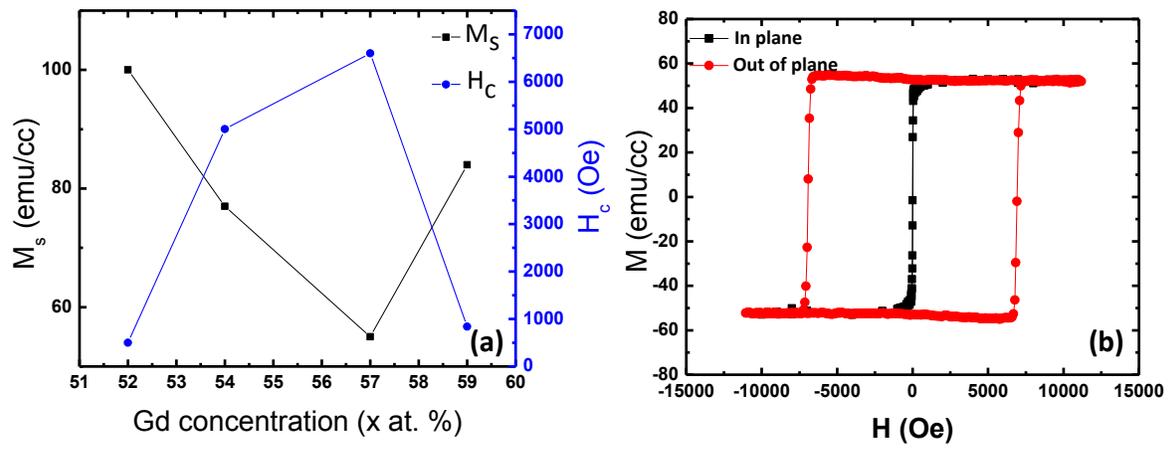